\documentclass[footinbib,a4paper,aps,prb,reprint,preprintnumbers,amsmath,amssymb, superscriptaddress]{revtex4-2}
\usepackage{verbatim}
\usepackage{color}
\usepackage{tikz}
\usepackage{xr}
\usepackage[colorlinks=true,citecolor=blue,linkcolor=blue,urlcolor=blue]{hyperref}
\usepackage{braket}
\usepackage[normalem]{ulem}
\usepackage{upgreek}
\usepackage[percent]{overpic}
\usepackage{graphicx,xcolor}
\usepackage{dcolumn}
\usepackage{bm}
\usepackage[shortlabels]{enumitem}
\usepackage{bbold}
\usepackage{array}
\usepackage{orcidlink}

\newcolumntype{C}[1]{>{\centering\arraybackslash}m{#1}}

\renewcommand{\rm}[1]{\textrm{#1}}

\newcommand{\mbf}[1]{\mathbf{#1}}

\renewcommand{\ket}[1]{|{#1}\rangle}
\renewcommand{\bra}[1]{\langle{#1}|}
\newcommand{\bracket}[2]{\langle{#1}|{#2}\rangle}

\definecolor{forestgreen}{rgb}{0.1, 0.6, 0.2}

\makeatletter
\renewcommand{\fnum@figure}{FIG. \thefigure}
\makeatother

\begin{document}

\newcolumntype{C}[1]{%
 >{\vbox to 50ex\bgroup\vfill\centering}%
 p{#1}%
 <{\egroup}}

\title{Calculating strongly correlated ground states from the non-Markovian dissipative dynamics of Gaussian fermions}
\author{Giuliano Chiriacò~\orcidlink{0000-0002-3906-4437}}
\email{giuliano.chiriaco@dfa.unict.it}
\affiliation{Dipartimento di Fisica e Astronomia, Università di Catania, Via Santa Sofia 64, 95123, Catania, Italy }

\begin{abstract}
We introduce a mapping between the ground state of interacting fermionic Hamiltonians and the non-equilibrium steady state of a purely dissipative open quantum system. Within the framework of third quantization, we map the Fermi-Hubbard Hamiltonian onto Lindblad jump operators acting on Majorana fermions. Remarkably, both hopping and interaction terms map onto jump operators that preserve the Gaussianity of Majorana fermions along individual quantum trajectories. As a result, the dynamics can be unravelled and each trajectory can be simulated efficiently using only two-point correlation functions, with a computational cost that scales polynomially with system size. We further show that finite particle number requires negative dissipative rates, leading to an intrinsically non-Markovian dynamics. The corresponding trajectory unravelling involves both positive and negative stochastic weights and exhibits a sign problem and large fluctuations, so that convergence requires an exponentially large number of trajectories. The overall computational cost remains exponential in system size despite the efficient Gaussian representation of individual trajectories, but is crucially dependent on the computational complexity of the non-Markovian unravelling, motivating further studies on the efficiency of such unravellings.
\end{abstract}

\maketitle

\section{Introduction}

Strongly correlated systems and open quantum systems are two major areas of interest in modern physics. Strongly correlated systems \cite{RMP_MIT, Khomskii_2014, BruusFlensberg} model most quantum materials, and the study of their ground state is crucial to understand many significant phenomena in condensed matter physics, such as superconductivity, magnetism, topological materials, metal-insulator transitions, etc. However, few correlated models are exactly solvable and exact numerical simulations are not feasible, since the Hilbert size of a many-body system scales exponentially with the size of the system, so one has to often rely on approximate numerical techniques (e.g. DMFT, DMRG, tensor networks) \cite{DMFT_Georges,DMFT_Kotliar,DMRG_White,DMRG_Schollwock,DMRG_Verstraete01032008,DMRG_Orus}. Meanwhile, open quantum systems \cite{Breuer:Petruccione,gardinerQuantumNoise2010,Rivas:Huelga} describe the properties of a system coupled to an environment, whose exact treatment is impractical. Such systems exhibit a rich phenomenology, including non-equilibrium and transient phases, measurement-induced criticality, and much more \cite{Basov11,Lee13,nakamuraElectricfieldinducedMetalMaintained2013,chiriacoVoltageinducedMetalinsulatorTransition2018b,chiriacoPolarityDependentHeating2020b,Mitrano16,Fausti11,chiriacoTransientSuperconductivitySuperconductivity2018b,Kogar20,chiriacoNegativeAbsoluteConductivity2020b,Li18,Li2019,Jian20,sierantDissipativeFloquetDynamics2022a,Sieberer13,Skinner2019,Jian21,Li20,Li21,ippolitiEntanglementPhaseTransitions2021,Turkeshi2021a,carolloDissipativeQuasiparticlePicture2022,albaFreeFermionsDephasing2023,albertonEntanglementTransitionMonitored2021,coppolaGrowthEntanglementEntropy2022,Romito_PhysRevA.110.022214_2024,Mirlin2023,Gribben2024,Kelly_2025,piccittoEntanglementDynamicsString2023,Piccitto_2025,RomitoPRX,Xing_PRL2026}. They play a central role in understanding noise and decoherence processes in quantum optics and quantum information, and their understanding is thus essential for the development of quantum technologies \cite{Preskill_2018,Paladino2014,Fazio}.

A strongly correlated system is characterized via a Hamiltonian operator acting on the quantum states of the system. An open quantum system (OQS) is typically described by a master equation  -- e.g. the Lindblad-Gorini-Kossakowski-Sudarshan equation, the Redfield equation, etc -- that determines its time evolution via the Liouvillian, a superoperator acting on the OQS density matrix. These two descriptions are connected by a one-to-one mapping between the density matrix of an OQS and the quantum states of a correlated system with twice the number of degrees of freedom, which relates the Lindbladian of an OQS to a Hamiltonian operator, and maps the non-equilibrium steady state (NESS) of the open quantum system to the ground state (GS) of the corresponding Hamiltonian \cite{prosenThirdQuantizationGeneral2008,seligmanThirdQuantization2010,Prosen_2010,mcdonaldThirdQuantizationOpen2023}. Recent works \cite{Prosen2008_XY,Prosen_2010XY,Prosen2011_XY} used this mapping to express the NESS of various dissipative systems subject to a Lindblad dynamics as the GS of an exactly solvable quadratic Hamiltonian.

This work takes a complementary perspective: given a generic interacting Hamiltonian, can its ground state be studied as the NESS of an appropriate master equation? The answer is yes. Here, we show that a hermitian Hamiltonian corresponds to a purely dissipative Lindblad dynamics with eternal non-Markovian terms. We focus on the Fermi-Hubbard model and map it to an OQS of many Majorana modes. We show that the density-density interaction terms (which are responsible for the computational complexity) map to a correlated dephasing dissipation acting on a system of many qubits. This result is not present in the literature \cite{Prosen2008_XY,Prosen_2010XY,Prosen2011_XY}, where only non-interacting, free-fermionic Hamiltonians and Lindbladians are considered; indeed, it was not guaranteed a priori that a quartic fermionic term could be mapped into a tractable dissipative term.

\begin{figure*}[t!]
    \centering
    \includegraphics[width=0.96\linewidth]{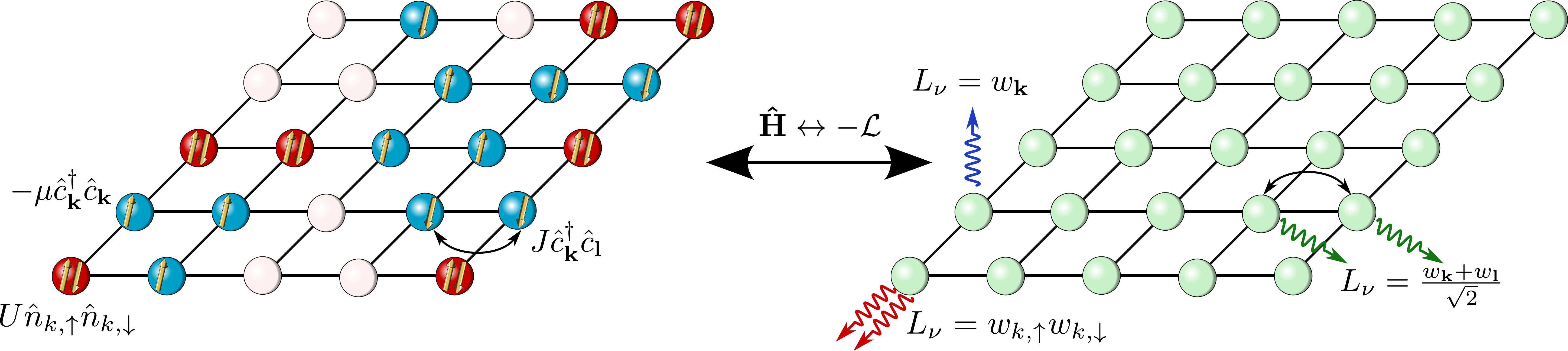}
    \caption{Sketch of the model considered and of the mapping. To the left we have the standard depiction of a Fermi-Hubbard model of electrons, with on-site repulsion $U$, hopping $J$ and chemical potential $\mu$. To the right we have the mapping to the dissipative open quantum systems of Majorana fermions, where each Hamiltonian term maps to a dephasing noise (wavy arrows). The $\mu$ term maps to a single site noise, the $J$ term maps to a superposition of dephasing noise on heighboring sites, the $U$ term maps to correlated dephasing in time.}
    \label{fig:drawing}
\end{figure*}

The dissipative dynamics can then be investigated using the quantum trajectory technique \cite{gardinerWavefunctionQuantumStochastic1992,plenioQuantumjumpApproachDissipative1998,molmerMonteCarloWavefunction1993,daleyQuantumTrajectoriesOpen2014a}, adapted for a non-Markovian dynamics \cite{breuerColloquiumNonMarkovianDynamics2016a,beckerQuantumTrajectoriesTimeLocal2023}. A remarkable feature of the quantum trajectories approach is that the resulting dynamics preserves the Gaussianity \cite{suraceFermionicGaussianStates2022,fagottiEntanglementEntropyTwo2010,ferraroGaussianStatesContinuous2005,braskGaussianStatesOperations2022,genoniQuantifyingNonGaussianCharacter2008,braunsteinQuantumInformationContinuous2005,chiriacoComputableMeasuresNonMarkovianity2025} of the Majorana modes along each trajectory, so that the time evolution can be fully captured using two-point correlation functions, a computational task that scales polynomially with system size. This is a remarkable result: interactions terms in the Hamiltonian map to a correlated dissipation which is still tractable in polynomial time via quantum trajectories. However, the unraveling of a non-Markovian master equation, which is an open problem subject to ongoing research \cite{GambettaNonMarkov,breuerGenuineQuantumTrajectories2004,piiloNonMarkovianQuantumJumps2008,piiloOpenSystemDynamics2009,BreuerPiiloEPL_2009_nonMarkov,breuerColloquiumNonMarkovianDynamics2016a,megierEternalNonMarkovianityRandom2017,smirneRateOperatorUnraveling2020,chiriacoDiagrammaticMethodManybody2023a,donvilQuantumTrajectoryFramework2022,beckerQuantumTrajectoriesTimeLocal2023,settimoGeneralizedrateoperatorQuantumJumps2024,settimo2026quantumjumpunravelingsnonmarkovian,tsitsishviliMeasurementInducedTransitions2024,muzziEntanglementEnhancementInduced2025}, exhibits a sign problem analogous to the one encountered in Quantum Monte Carlo (QMC) simulations \cite{QMC_White,QMC_Wiese,QMC_Trebst,Pan_2024}. In particular, it appears that the number of trajectories needed to simulate the master equation dynamics scales exponentially with system size.

Therefore, it seems that while the computational complexity of simulating one trajectory is reduced to a polynomial scaling with system size, the exponential complexity reappears in the number of trajectories. While the mapping developed in this work does not eliminate the exponential complexity, our results offer a new perspective on the complexity of finding the ground state of correlated systems, and open several directions for future research, in particular regarding the sign problem and the unravelling of non-Markovian master equations.

The paper is organized as follows. In Section \ref{sec:model} we present the notation and the model of our system and in \ref{sec:mapping} we describe the mapping between the correlated system and the open quantum system. In Section \ref{sec:traj} we describe the unravelling protocol used to evolve the open quantum system along the trajectories. In Section \ref{sec:Gaussian} we show how to employ correlation functions of Gaussian states to describe the trajectory dynamics and calculate observables. In Section \ref{sec:complexity} we analyze the computational cost of this method and discuss the sign problem associated to the non-Markovian unravelling. Finally in Section \ref{Sec:Conclusions} we summarize the results of this work and present our conclusions.

\section{The model}\label{sec:model}

We consider a system of strongly correlated fermions with spin $\frac12$ (e.g. electrons) defined on a lattice with $L$ sites identified by the label $k$. The bold font label $\mbf k\equiv(k,\sigma)$ identifies an electron on site $k$ with spin $\sigma=\uparrow,\downarrow$. Operators acting on the fermions are identified with a hat, e.g. $\hat c_{\mbf k}$ destroys a fermion in the state $\mbf k$. The quantum state of the fermions is described via a ket notation, e.g. $\ket\psi$.

This system can be mapped to a system of Majorana fermions on $L$ sites, which consists of $2L$ Majorana modes and is equivalent to a system of $L$ qubits via a Jordan-Wigner transformation). The state of the Majorana fermions is described by a density matrix $\rho$ and by the $2L$ Majorana operators $w_{\mbf k}$ where again $\mbf k=(k,\sigma)$, such that $\{w_{\mbf{k}},w_{\mbf {l}}\}=2\delta_{\mbf k,\mbf l}$ \footnote{the Majorana fermions are related to spinless fermions via $w_{k,\uparrow}\equiv a_k+a_k^{\dagger}$, $w_{k,\downarrow}\equiv i(a_k-a_k^{\dagger})$, with $a_k^{\dagger}$ and $a_k$ the fermionic creation and annihilation operators.}. In the canonical basis, the Majorana operators have the expression
\begin{equation}\label{eq:MajoranaOp}
w_{k,\uparrow}=\begin{pmatrix}0&1\\1&0\end{pmatrix}_k;\qquad w_{k,\downarrow}=\begin{pmatrix}0&-i\\i&0\end{pmatrix}_k.
\end{equation}

The strongly correlated fermions are described by a generic Hamiltonian $\mbf{\hat H}$, of which we wish to find the ground state $\ket{\psi_{gs}}$ such that $\mbf{\hat H}\ket{\psi_{gs}}=E_{gs}\ket{\psi_{gs}}$. 

We employ the third quantization technique in reverse: we map the correlated fermions to a system of $L$ Majorana fermions described by a density matrix $\rho$. The Hamiltonian maps to a Lindblad superoperator $\mathcal{L}$: $\mbf{\hat H}\leftrightarrow-\mathcal L$ and $\mbf{\hat H}\ket{\psi}\leftrightarrow-\mathcal{L}\rho$. The ground state energy (lowest eigenvalue of $\mbf{\hat H}$) is mapped to the largest eigenvalue of $\mathcal{L}$ and the ground state $\ket{\psi_{gs}}$ is mapped to the corresponding eigenvector $\rho_{\infty}$: $\mathcal{L}\rho_{\infty}=-E_{gs}\rho_{\infty}$. Formally $\rho_{\infty}$ can be found as the infinite time limit of the master equation $\partial_t\rho=\mathcal{L}\rho$:
\begin{equation}\label{Eq:EquivalenceGS-SS}
\ket{\psi_{gs}}\,\,\leftrightarrow\,\,\rho_{\infty}=\lim_{t\rightarrow\infty}e^{\mathcal{L}t}\rho(t=0).
\end{equation}

This framework has been used \cite{prosenThirdQuantizationGeneral2008,mcdonaldThirdQuantizationOpen2023} to find the steady state of a quadratic Lindbladian as the ground state of a quadratic non-Hermitian Hamiltonian. Here we are interested in the opposite question: how do we find the ground state of a correlated Hamiltonian by mapping it to a suitable Lindbladian dynamics?

We investigate this question starting from a generalization of the Fermi-Hubbard model
\begin{equation}\label{Eq:FH_Ham}
\mbf{\hat {H}}= \sum_{\bf k\bf l}J_{\bf k\bf l}(\hat c^{\dagger}_{\bf k}\hat c_{\bf l}+\text{h.c.}) + \sum_{\bf k,\bf l}U_{\bf k\bf l}\hat n_{\bf k}\hat n_{\bf l}-\mu\sum_{\bf k}\hat c^{\dagger}_{\bf k}\hat c_{\bf k},
\end{equation}
where $\hat{c}^\dagger_{\mbf k}$/$\hat{c}_{\mbf k}$ creates/destroys a fermion on site $k$ with spin $\sigma$ and $\hat n_{\mbf k}\equiv \hat{c}^\dagger_{\mbf k}\hat{c}_{\mbf k}$ is the fermion density in state $\mbf k$. The term $U$ describes density-density interactions (repulsive for $U>0$), $J$ describes the hopping terms (we have absorbed the conventional minus sign into the definition of $J_{\mbf{kl}}$), and $\mu$ is the chemical potential. The standard Fermi-Hubbard model is recovered when $U_{(k,\sigma)(l,\sigma')}=\frac U2\delta_{k,l}\delta_{\sigma,-\sigma'}$, describing on-site repulsion between fermions with different spin, and $J_{(k,\sigma)(l,\sigma')}=J\delta_{\langle{kl}\rangle}\delta_{\sigma,\sigma'}$, describing hopping between next neighbor sites $\langle kl\rangle$. We assume that $\mu$ is chosen so that the number of fermions in the system is even, in order to avoid parity issues when mapping to an OQS \cite{prosenThirdQuantizationGeneral2008}.

The Hamiltonian in Eq. \eqref{Eq:FH_Ham}  can be rewritten as
\begin{gather}\label{Eq:H2}
\hat H=\sum_{\bf k,\bf l}(\hat h_{J,\bf k\bf l}+\hat h_{U,\bf k\bf l})+\sum_{\bf k}h_{\mu,\mbf k};\\
\notag\hat h_{U,\bf k\bf l}=U_{\bf k\bf l}\hat n_{\bf k}\hat n_{\bf l}-\frac {U_{\bf k\bf l}}2(\hat n_{\bf k}+\hat n_{\bf l}); \quad h_{\mu,\mbf k}=-\mu_{\bf k}\hat c^{\dagger}_{\bf k}\hat c_{\bf k};\\
\notag\hat h_{J,\bf k\bf l}=J_{\bf k\bf l}(\hat c^{\dagger}_{\bf k}+\hat c^{\dagger}_{\bf l})(\hat c_{\bf k}+\hat c_{\bf l});\quad\mu_{\bf k}=\mu+\sum_{\bf l}(2J_{\bf k\bf l}-U_{\bf k\bf l}),
\end{gather}
where each $\hat h$ describes a different term in the Hamiltonian.

\begin{table}[b]
    \centering
    \begin{tabular}{|c|c|c|c|}
            \hline
        Term & $\hat h_{\nu}$ & $L_{\nu}$ & $\gamma_{\nu}$\\[0.5ex]\hline
        Hopping & $J_{\bf k\bf l}(\hat c^{\dagger}_{\bf k}+\hat c^{\dagger}_{\bf l})(\hat c_{\bf k}+\hat c_{\bf l})$& ${\color{white} \Big(}\frac{w_{\bf k}+w_{\bf l}}{\sqrt2}{\color{white} \Big)}$ & $J_{\bf k\bf l}$\\[1ex]
        1-body & $-\mu_{\bf k}\hat c^{\dagger}_{\bf k}\hat c_{\bf k}$& $w_{\bf k}$ & $-\frac{\mu_{\bf k}}2$\\[0.9ex]
        2-body & $U_{\bf k\bf l}\hat n_{\bf k}\hat n_{\bf l}-\frac {U_{\bf k\bf l}}2(\hat n_{\bf k}+\hat n_{\bf l})$& $w_{\bf k}w_{\bf l}$ & $-\frac {U_{\bf k\bf l}}4$\\[1ex]
            \hline
    \end{tabular}
    \caption{Mapping between a Hamiltonian term (hopping, single-body or two-body interactions) and the corresponding jump operator and dissipative rate in the Lindbladian formulation.}
    \label{tab:map}
\end{table}

\subsection{Mapping rules}\label{sec:mapping}

We now map our system to an open quantum system (OQS) of Majorana fermions via the third quantization technique, see Fig.\ref{fig:drawing}. The OQS is described by Majorana fermion operators $w_{\bf k}$ (with ${\bf{k}}=(k,\sigma)$). Within this mapping, the empty state $\ket{0}$ maps to the identity density matrix $\rho=\mathbb{1}$, $\ket{\uparrow_k}=c_{k,\uparrow}^\dagger\ket{0}$ maps to $\rho=w_{k,\uparrow}$, and $\ket{\downarrow_k}=c_{k,\downarrow}^\dagger\ket{0}$ maps to $\rho=w_{k,\downarrow}=\begin{pmatrix}0&-i\\i&0\end{pmatrix}_k$. In general, a generic state $\ket{\psi_\rho}$ maps to a density matrix describing the state of the OQS
\begin{equation}\label{eq:psi_rho_mao}
    \ket{\psi_\rho}=\prod_{\mbf k}(c_{\mbf k}^{\dagger})^{\alpha_{\mbf k}}\ket{0}\,\,\leftrightarrow\,\,\rho=\prod_{\mbf k}w_{\mbf k}^{\alpha_{\mbf k}}, \quad \alpha_{\bf k}=0,1.
\end{equation}

The Hamiltonian $\hat{\bf H}$ maps to a Lindbladian $\hat{\bf H}\,\,\leftrightarrow\,\,-\mathcal{L}$ such that
\begin{gather}\label{Eq:Lind}
\dot\rho=\mathcal{L}\rho;\qquad \mathcal{L}=\sum_{\nu}\gamma_\nu\mathcal{L}[L_\nu];\\
\notag\mathcal{L}[L_\nu](\rho) \equiv  L_\nu \rho L_\nu^\dagger - \frac{1}{2} \{L_\nu^\dagger L_\nu, \rho\},
\end{gather}
where each $\mathcal{L}_\nu$ corresponds to a specific $\hat h$ in Eq. \eqref{Eq:H2} as described in Table \ref{tab:map}. The dependence $L_\nu(w_\mbf k)$ can be found applying the rules of the mapping, see Appendix \ref{app:Mapping}. For example
\begin{equation}\label{eq:MappingExample}
\hat c^\dagger_\mbf k\hat c_\mbf k\ket{\psi_\rho}\,\leftrightarrow\,\frac12\rho-\frac12w_\mbf k\rho w_\mbf k;\quad\hat h_{\mu,\mbf k}\,\leftrightarrow\,-\frac{\mu_\mbf k}{2}\mathcal{L}[w_\mbf k].
\end{equation}

Expectation values can be calculated using the mapping of the scalar product $\bracket{\psi_\rho}{\psi'_\rho}\leftrightarrow\frac{\text{Tr}(\rho^\dagger\rho')}{||\rho||\cdot||\rho'||}$ where $||\rho||\equiv\sqrt{\text{Tr}(\rho^\dagger\rho)}$, so that
\begin{equation}\label{eq:Oexp_map}
    \langle \hat{\mbf O}\rangle\equiv\bra{\psi_{gs}}\hat{\mbf O}(\hat c^\dagger_{\mbf k},\hat c_{\mbf k})\ket{\psi_{gs}}\leftrightarrow\frac{\text{Tr}(\rho_{\infty}^\dagger {O}[w_{\mbf k}]\rho_{\infty})}{\text{Tr}(\rho_{\infty}^\dagger\rho_{\infty})},
\end{equation}
where $\hat{\mbf O}$ is an observable expressed in terms of the fermionic operators and $O$ is its mapping in terms of the Majorana. The exact rules of the mapping are reported in Appendix; here we show the example of the density $\hat{\mbf O}=\hat c^\dagger_{\mbf k}\hat c_{\mbf k}=\hat n_{\mbf k}$:
\begin{equation}\label{eq:ExpValDen}
\langle\hat n_{\mbf k}\rangle=\frac12\frac{\text{Tr}[\rho_{\infty}^\dagger(\rho_{\infty}-w_{\mbf k}\rho_{\infty}w_{\mbf k})]}{\text{Tr}(\rho_{\infty}^\dagger\rho_{\infty})}.
\end{equation}

We note that the hopping and chemical potential terms in $\hat {\mbf H}$ map to $L_\nu$ that are linear in the Majorana operators; this result was crucial in Ref. \cite{prosenThirdQuantizationGeneral2008} where dissipative dynamics with only linear jump operators was studied via the mapping to an exactly solvable Hamiltonian. On the other hand, density-density interaction terms map into jump operators that are quadratic in the Majorana operators. We observe that all jump operators are Hermitian, so that they induce dephasing-like terms: $\hat h_{\mu,\mbf k}$ maps to a local dephasing term, $\hat h_{J,\mbf k\mbf l}$ maps to space-correlated dephasing between $\mbf k$ and $\mbf l$, while $\hat h_{U,\mbf k\mbf l}$ maps to time-correlated dephasing due to $L_\nu=w_\mbf kw_\mbf l$. It is interesting that correlations in space in the interacting fermionic systems map to correlations in time in the OQS.

We point out that positive values of $\mu_{\mbf k}$ are required to have a non-zero number of particles in the correlated system, resulting in a negative dissipation rate $\gamma=-\frac{\mu}{2}<0$. Thus the master equation \eqref{Eq:Lind} is a time-local eternal non-Markovian equation, since dissipation rates are constant in time, but at least one of them is negative \footnote{If all dissipation rates $\gamma_\nu$ were positive (necessary condition to have a Markovian master equation) the NESS would be the trivial identity matrix, corresponding to an infinite temperature state -- as expected for a Markovian master equation with dephasing jump operators.}. We note that for a non-Markovian dynamics it is not technically correct to talk about a steady state, since quantum coherences could grow exponentially instead of decaying; however, this is not an issue both conceptually and numerically since the normalization factor $\text{Tr}(\rho^\dagger_\infty\rho_\infty)$ in Eq. \eqref{eq:Oexp_map} takes care of this behavior and ensures convergence of any expectation value. For this reason we continue to refer to $\rho_\infty$ as NESS.

\section{Dynamics via quantum trajectories}\label{sec:traj}

The results of the previous section show that finding the ground state of a correlated system (whose Hilbert space dimension scales as $2^{2L}$) is equivalent to determining the non-equilibrium steady state (NESS) of a purely dissipative dynamics. However, solving for the NESS by directly integrating the Lindblad equation is still an exponentially complex task in the size of the system ($\sim2^{2L}$).

A more efficient alternative is to employ the Monte Carlo Quantum Jumps (MCQJ) method, which unravels the Lindblad dynamics into stochastic quantum trajectories \cite{daleyQuantumTrajectoriesOpen2014a}. Instead of evolving the full density matrix $\rho$, the MCQJ method tracks the evolution of pure quantum states along a trajectory $a$, $\rho_{a}=\ket{\psi_{a}}\bra{\psi_{a}}$, subject to the random application of the jump operators appearing in the master equation. The probability of applying each jump operator is proportional to the corresponding dissipative rate. The solution to the Lindblad equation is then recovered by averaging over a statistically significant number $N_{\rm tr}$ of trajectories
\begin{equation}\label{eq:QTrajMark}
\rho(t)=\lim_{N_{\rm tr}\rightarrow\infty}\frac1{N_{\rm tr}}\sum_{a=1}^{N_{\rm tr}}\rho_a(t).
\end{equation}

The MCQJ method is well established for Markovian Lindblad master equations, and has found extensive applications. However, its application to non-Markovian master equations is less straightforward. The main issue is that a negative dissipation rate would lead to a negative jump probability. Several methods have been proposed to deal with this problem \cite{breuerGenuineQuantumTrajectories2004,piiloNonMarkovianQuantumJumps2008,piiloOpenSystemDynamics2009,BreuerPiiloEPL_2009_nonMarkov,breuerColloquiumNonMarkovianDynamics2016a,megierEternalNonMarkovianityRandom2017,smirneRateOperatorUnraveling2020,chiriacoDiagrammaticMethodManybody2023a,donvilQuantumTrajectoryFramework2022,beckerQuantumTrajectoriesTimeLocal2023,settimoGeneralizedrateoperatorQuantumJumps2024,settimo2026quantumjumpunravelingsnonmarkovian}. One method is to introduce a classical bit $s_a(t)$ (whose initial value is $s_a(0)=1$) which tracks the trace of $\rho_a$ and whose sign changes every time a jump associated to a negative $\gamma_\nu$ occurs \cite{beckerQuantumTrajectoriesTimeLocal2023}.

Within this framework, the evolution of a pure state is governed by the following stochastic protocol. For each time step $\delta t$ the system undergoes either a \textit{quantum jump} or a \textit{non-Hermitian evolution}.

\textit{Quantum Jump} occurs with operator $L_\nu$ with probability $p_\nu = |\gamma_\nu| dt$. The quantum state and the classical variable evolve as 
\begin{equation}\label{eq:QJjump}
  \begin{aligned}
    \rho_{a}(t+dt)&=L_{\nu}\rho_{a}(t)L_{\nu}^{\dagger}; \\
         s_a(t+dt)&=\text{sign}(\gamma_\nu)s_a(t),
  \end{aligned}
\end{equation}
i.e. $s_a$ changes sign if a non-Markovian jump occurs. We note that the application of the jump operators conserves the trace of $\rho_a$ since $\text{Tr}(L_\nu\rho_a L^\dagger_\nu)=\text{Tr}(L^\dagger_\nu L_\nu\rho_a)=\text{Tr}\rho_a$; in fact, the Majorana operators are Hermitian and satisfy $w_{\bf k}^2=1$, and all the jump operators considered here (see Tab. \eqref{tab:map}) are also Hermitian $L_{\nu}=L_{\nu}^{\dagger}$ and satisfy $L_{\nu}^{\dagger}L_{\nu}=\mathbb1$

\textit{Non-Hermitian evolution} occurs with probability $1 - \sum_\nu p_\nu = 1 - \sum_\nu |\gamma_\nu| dt$. In standard MCQJ protocols (even extended to non-Markovian dynamics \cite{beckerQuantumTrajectoriesTimeLocal2023}) the state evolves with an effective Hamiltonian $H_{\rm{eff}}=-\frac i2\sum_{\nu}\gamma_{\nu}L_{\nu}^{\dagger}L_{\nu}=-\frac i2\sum_\nu\gamma_\nu$. Therefore the non-Hermitian evolution simply rescales the density matrix; within the framework of standard MCQJ protocols, the rescaling factor is
\begin{align}\label{eq:QJnonHerm}
    \rho_a(t+dt)=&\frac{(\mathbb1-iH_{\rm{eff}}dt)\rho_a(t)(\mathbb1+iH_{\rm{eff}}^\dagger dt)}{1-\sum_\nu|\gamma_\nu|dt}=\\
\notag=&\frac{1-\sum_{\nu}\gamma_{\nu}dt}{1-\sum_\nu|\gamma_\nu|dt}\rho_a(t).
\end{align}
This protocol was also used in Ref. \cite{beckerQuantumTrajectoriesTimeLocal2023} to unravel an eternal non-Markovian dynamics.

For a Markovian dynamics all $\gamma_\nu$ are positive so that the rescaling factor simplifies to 1, and the non-Hermitian evolution correctly conserves the trace. However, for non-Markovian dynamics some of the $\gamma_\nu$ are negative and the rescaling factor is strictly $>1$, leading to a growth of the trace of $\rho_a$. It is more convenient to choose a protocol that conserves the trace of $\rho_a$ and keeps track of the rescaling factor in the dynamics of $s_a$:
\begin{equation}\label{eq:QJnonH}
\rho_a(t+dt)=\rho_a(t); \quad s_a(t+dt)=\frac{1-\sum_{\nu}\gamma_{\nu}dt}{1-\sum_\nu|\gamma_\nu|dt}s_a(t).
\end{equation}

The solution of the master equation is then recovered as a weighted average with $s_a$ over the trajectories.
\begin{equation}\label{Eq:rho_avg_s}
\rho(t)\approx\frac1{N_{\rm{tr}}}\sum_{a=1}^{N_{\rm tr}}s_a(t)\rho_a(t).
\end{equation}

Our protocol is such that $\text{Tr}\rho_a(t)=1$ along each trajectory and $\text{Tr}\rho(t)=1$, while the absolute value of $s_a(t)$ can change over time. This is an alternative protocol to the one used in Ref. \cite{beckerQuantumTrajectoriesTimeLocal2023}, where the absolute value of $s_a$ is constant, but $\text{Tr}\rho_a(t)=1$ changes over time. We introduce this different protocol in order to enforce trace conservation, so that $\rho_a$ describes a physical state at any time along any trajectory. This is crucial to employ the Gaussian state formalism (introduced in the next Section), which could not be used with the protocol of Ref. \cite{beckerQuantumTrajectoriesTimeLocal2023} since it produces not-normalized states. In Appendix \ref{App:BitEvol} we prove that $\text{Tr}\rho(t)=1$.

\section{Gaussian evolution}\label{sec:Gaussian}

A crucial observation is that all jump operators in Table \eqref{tab:map} preserve the Gaussianity of a quantum state. A Gaussian fermionic state can be written as \cite{cheongManyBodyDensityMatrices2003,Chung_2001,peschelCalculationReducedDensity2003,peschelReducedDensityMatrices2009,fagottiEntanglementEntropyTwo2010,suraceFermionicGaussianStates2022}
\begin{equation}\label{eq:rhoGauss}
\rho=\frac1{Z(A)}e^{\frac14\sum_{\bf k\bf l}A_{\bf k\bf l}w_{\bf k}w_{\bf l}}\equiv \frac1{Z(A)}e^{A_{\bf k\bf l}w_{\bf k}w_{\bf l}/4},
\end{equation}
where the sum over repeated indices is omitted for book-keeping reasons. Here $A_{\mbf k \mbf l}$ is an antisymmetric matrix related to the covariance matrix $\Gamma_{\bf k\bf l}=\langle w_{\mbf k}w_{\mbf l}\rangle-\delta_{\mbf k\mbf l}=\tanh(A_{\bf k\bf l}/2)$, and $Z(A)\equiv\text{Tr}(e^{A_{\bf k\bf l}w_{\bf k}w_{\bf l}/4})$ is the so-called partition function. It can be proved that $\ln Z(A)=\frac12\text{Tr}\ln[2\cosh(A/2)]=-\frac12\text{Tr}\ln[(1+\Gamma)/2]$ \cite{chiriacoComputableMeasuresNonMarkovianity2025}.

It is possible to prove (see also Appendix \ref{app:UpdateCorr}) that applying any of the jump operators to a Gaussian state, the state remains Gaussian. If the state at time $t$ is described by $A_{\mbf k \mbf l}(t)$, then also the state at $t+dt$ is Gaussian 
\begin{equation}
\rho(t+dt)=\frac1{Z(A(t))}e^{\frac14A_{\bf k\bf l}(t+dt)w_{\bf k}w_{\bf l}}.
\end{equation}

We show this explicitly for a jump given by $w_\mbf j$:
\begin{gather}
\rho(t+dt)=w_\mbf j\frac{e^{\frac14A_{\bf k\bf l}(t)w_{\bf k}w_{\bf l}}}{Z(A)}w_\mbf j=\frac{e^{\frac14A_{\bf k\bf l}(t)w_\mbf jw_{\bf k}w_{\bf l}w_\mbf j}}{Z(A)},
\end{gather}
where we have used that $w_\mbf j$ is a unitary operator and can be moved to the exponent. By making use of the anticommutation rules for the Majorana operators, it is possible to reduce the exponent to a quadratic form $w_\mbf jw_{\bf k}w_{\bf l}w_\mbf j=w_{\mbf k}w_{\mbf l}(1-2\delta_{\mbf k\mbf j}-2\delta_{\mbf l\mbf j}+4\delta_{\mbf k\mbf j}\delta_{\mbf l\mbf j})$, so that this can be recast in Gaussian form with a new matrix $A_{\mbf k\mbf l}(t+dt)=A_{\mbf k\mbf l}(t)(1-2\delta_{\mbf k\mbf j}-2\delta_{\mbf l\mbf j}+4\delta_{\mbf k\mbf j}\delta_{\mbf l\mbf j})=O^{-1}_{w_\mbf j}A_{\mbf k\mbf l}O_{w_\mbf j}$ where $O_{w_\mbf j}$ is the matrix generating the unitary transformation of $A$ due to the application of $w_\mbf j$.

Indeed the action of any jump on $A$ can be expressed in terms of an appropriate unitary transformation. The rules to calculate the updated matrix $A'_{\mbf k\mbf l}=A_{\mbf k\mbf l}(t+dt)$ for the other jumps are given in Table \eqref{tab:A_update}. We note that we used standard techniques for Gaussian states to derive the update rules for $A$ for the Lindblad jump operators considered here, while existing works in the literature derive the update rules for $\Gamma$ \cite{coppolaGrowthEntanglementEntropy2022,tsitsishviliMeasurementInducedTransitions2024,muzziEntanglementEnhancementInduced2025}. We also remark that the computational complexity of each update operation is $\sim\mathcal{O}(L)$ instead of $\sim\mathcal{O}(L^2)$ since it requires the update of just one row and one columns of $A$, instead of the entire matrix.
\begin{table}[t]
    \centering
    \begin{tabular}{|c|c|}
        \hline
        ${\color{white} \Big(}L_{\nu}{\color{white} \Big)}$ & $A_{\bf k\bf l}(t+dt)$ \\[0.5ex]
        \hline

        ${\color{white} \Big(}w_{\mbf j}{\color{white} \Big)}$ &
        $\displaystyle A'_{\mbf k\mbf j}=-A_{\mbf k\mbf j},\,
        A'_{\mbf j\mbf l}=-A_{\mbf j\mbf l},\,
        A'_{\mbf j\mbf j}=A_{\mbf j\mbf j}$ \\[1ex]

        $\displaystyle \frac{w_{\bf i}+w_{\bf j}}{\sqrt2}$ &
        \parbox[c]{6.5cm}{\centering$\displaystyle A'_{\bf k\bf i}=-A_{\bf k\bf j},\quad
        A'_{\bf k\bf j}=-A_{\bf k\bf i},\quad   
        A'_{\bf i\bf l}=-A_{\bf j\bf l},$ $\qquad$ 
        $A'_{\bf j\bf l}=-A_{\bf i\bf l},\quad
        A'_{\mbf i\mbf j}=A_{\mbf i\mbf j}$}\\[1ex]
        \hline
    \end{tabular}
    \caption{Rules for the update of the matrix $A$ when applying different jump operators. The application of $L_\nu=w_{\mbf k}w_{\mbf l}$ simply results in the consecutive application of $w_{\mbf l}$ and then $w_{\mbf k}$.}
    \label{tab:A_update}
\end{table}

The conserved Gaussian structure implies that the state is fully described by its covariance matrix at any time along a quantum trajectory. In particular, since the NESS in Eq. \eqref{Eq:EquivalenceGS-SS} does not depend on the initial state of $\rho$, it is possible to employ the MCQJ method starting from a Gaussian state, and express the average density matrix and any expectation value as an average over trajectories
\begin{gather}\label{eq:QTrajnonMark}
\rho(t)\approx\frac1{N_{\rm{tr}}}\sum_{a=1}^{N_{\rm tr}}s_a(t)\frac{e^{\frac14A^a_{\bf k\bf l}(t)w_{\bf k}w_{\bf l}}}{Z(A^a(t))};\\
\notag\text{Tr}(\rho_\infty^\dagger\rho_\infty)=\frac1{N_\rm{tr}^2}\sum_{a,b}s_as_b\frac{\text{Tr}(e^{\frac14A^a_{\bf k\bf l}w_{\bf k}w_{\bf l}}e^{\frac14A^b_{\bf k\bf l}w_{\bf k}w_{\bf l}})}{Z(A^a)Z(A^b)}.
\end{gather}

It is possible to show \cite{fagottiEntanglementEntropyTwo2010, chiriacoComputableMeasuresNonMarkovianity2025} that the product of two Gaussian states $\rho_a$and $\rho_b$ is still in Gaussian form, although it generally does not represent a physical state, unless $\rho_a$ and $\rho_b$ commute. This is not an issue, since we are only interested in the overlap $\text{Tr}(\rho_a^\dagger\rho_b)$ and do not need it to correspond to a physical state. We find that
\begin{equation}\label{eq:Aplus}
e^{\frac14A^a_{\bf k\bf l}w_{\bf k}w_{\bf l}}e^{\frac14A^b_{\bf k\bf l}w_{\bf k}w_{\bf l}}=e^{\frac14A^c_{\bf k\bf l}w_{\bf k}w_{\bf l}}; \quad e^{A^c}=e^{A^a}e^{A^b},
\end{equation}
so that $A^c\equiv A^a\oplus A^b$ is obtained via the matrix multiplication of the matrix exponentials. Note that the $\oplus$ appearing in the definition of $A^c$ is not the ordinary matrix addition, since in general $A^a$ and $A^b$ do not commute and $A^c$ does not need to equal $A^a+ A^b$. Thus $\text{Tr}(\rho_\infty^\dagger\rho_\infty)=\frac1{N_\rm{tr}^2}\sum_{a,b}s_as_b\frac{Z(A^a\oplus A^b)}{Z(A^a)Z(A^b)}$ reduces to the calculation of partition functions over trajectories. Expectation values of an operator $\hat{\mbf O}$ depend on
\begin{equation}\label{eq:operator_numerator}
    \text{Tr}(e^{\frac14A^a_{\bf k\bf l}w_{\bf k}w_{\bf l}}O[w_{\mbf j}]e^{\frac14A^b_{\bf k\bf l}w_{\bf k}w_{\bf l}})=Z(A^a\oplus A'^b),
\end{equation}
where $A'^b=O^{-1}A^b$ is the associated transformation of the matrix $A^b$. Thus, we write the expectation value as
\begin{gather}
\label{eq:OperatorGauss}
\langle\hat{\mbf O}\rangle=\frac{\sum_{a,b}s_as_b\frac{Z(A^a\oplus A'^b)}{Z(A^a)Z(A^b)}}{\sum_{a,b}s_as_b\frac{Z(A^a\oplus A^b)}{Z(A^a)Z(A^b)}};\\
\label{eq:OperatorGaussGamma}\langle\hat{\mbf O}\rangle=\frac{\sum_{a,b}s_as_b\sqrt{\text{det}\frac{1+\Gamma_a\Gamma_b'}{2}}}{\sum_{a,b}s_as_b\sqrt{\text{det}\frac{1+\Gamma_a\Gamma_b}{2}}},
\end{gather}
where Eq. \eqref{eq:OperatorGaussGamma} can be found by expressing $A$ in terms of the associated correlation matrices \cite{chiriacoComputableMeasuresNonMarkovianity2025}.

In summary, we can find the expectation value of an operator $\hat{\mbf O}$ via the following protocol:
\begin{itemize}
    \item we run the quantum jumps dynamics for an appropriate number of trajectories $N_{\rm{tr}}$ following the protocol described by Eqs. \eqref{eq:QJjump}-\eqref{eq:QJnonHerm} and Tab. \eqref{tab:A_update}, until convergence in time is reached. For each trajectory we obtain a classical bit $s_a$ and a matrix $A_a$.
    \item We find the action of $\hat{\mbf O}$ on the matrix $A$ following the rules of Tabs. \eqref{tab:map}-\eqref{tab:A_update}, and calculate $A'^b=O^{-1}A^bO$ for each trajectory.
    \item We calculate the associated correlation matrices $\Gamma_a=\tanh(A^a/2)$ and $\Gamma_b'=\tanh(A'^b/2)$ and calculate the weighted averages following Eq. \eqref{eq:OperatorGaussGamma}.
\end{itemize}  

We note that is not required that the final state of the evolution $\rho_\infty$ is generally not a Gaussian state, since it is the average over many Gaussian states. The only requirement is for the initial Gaussian state of each trajectory to have a non zero overlap with the steady state, as to avoid convergence issues; this can be obtained by choosing a random initial Gaussian state for each trajectory.

\section{Computational cost}\label{sec:complexity}

In this section, we estimate the computational complexity $\mathcal{C}$ required by the various numerical approaches.

A correlated system with $L$ sites, has a Hilbert space dimension $2^{2L}$, so that we can expect the computational complexity of any numerical algorithm that calculates the ground state properties to scale exponentially $\mathcal{C}\sim2^{2L}$.

On the other hand, the Gaussian trajectories method scales linearly with the number of trajectories, and the computational cost of each trajectory is polynomial in $L$. In fact, the cost of updating $A$, i.e. simulating one quantum jump, is $\mathcal{O}(L)$, while the number of quantum jumps occurring each time step is $\sim L$, so that the cost of simulating the evolution of $A(t)$ scales polynomially as $L^2$, as opposed to the exponential scaling in the Hilbert space dimension. The average in Eq. \eqref{eq:OperatorGaussGamma} requires diagonalizing $L\times L$ matrices for $N_{\rm{tr}}^2$ times, a task whose complexity scales as $\sim N_{\rm{tr}}^2L^2$, which is larger than the complexity $\sim N_{\rm{tr}}L^2$ of the evolution part of the protocol. Thus, by exploiting the Gaussianity-preserving nature of the dissipative dynamics, the complexity of the trajectories protocol is $\mathcal{C}\sim L^2N_{\rm{tr}}^2(L)$.

We point out that the dependence of $N_{\rm{tr}}$ with $L$ is crucial. For standard Markovian quantum trajectories, typically $N_{\rm{tr}}$ does not depend on $L$ and good convergence is already reached for $N_{\rm{tr}}\gtrsim100$, since fluctuations on the trajectory average scale as $1/\sqrt{N_{\rm{tr}}}$ \cite{daleyQuantumTrajectoriesOpen2014a}.

However, for a non-Markovian unravelling the question is more complicated, as also pointed out in \cite{beckerQuantumTrajectoriesTimeLocal2023}. The reason is that fluctuations of the classical bit $s(t)$ grow exponentially with time. In particular, we find that (see Appendix \ref{App:BitEvol})
\begin{gather}
\label{eq:st}
\overline{s(t+dt)}=\overline{s(t)}\\
\label{eq:s2t}
\overline{s^2(t+dt)}=\overline{s^2(t)}+dt\sum_{\gamma_\nu<0}2|\gamma_\nu|\overline{s^2(t)}
\end{gather}

Therefore while $\overline{s(t)}=1$, $\overline{s^2(t)}\sim e^{2\gamma_At}$, with $\gamma_A\equiv\sum_{\gamma_\nu<0}|\gamma_\nu|$. Thus fluctuations grow exponentially in time: $\Delta s=\sqrt{\overline{s^2}-\overline{s}^2}\sim e^{\gamma_At}$ while the average stays constant. This is a feature of any unravelling with negative rates $s_a(t)$, which significantly slows down convergence due to the oscillations between positive and negative values of $s_a(t)$. This is a similar issue to the sign problem in Quantum Monte Carlo (QMC) simulations, where negative weights in the statistical sampling complicate numerical convergence \cite{QMC_White,QMC_Wiese,Pan_2024,QMC_signBounds}.

The fluctuations on the average value of $s$, which determine the convergence of any expectation value, are thus $\Delta\overline{s}=\Delta s/\sqrt{N_{\rm{tr}}}\sim e^{\gamma_A t}/\sqrt{N_{\rm{tr}}}$. As time passes, fluctuations grow larger and a larger number of trajectories is needed to achieve convergence. If typical expectation values relax to their steady state value with rate $\gamma_R$, then trajectories need to be simulated for a time $\sim1/\gamma_R$ and fluctuations grow up to $\Delta\overline{s}\sim e^{\gamma_A/\gamma_R}/\sqrt{N_{\rm{tr}}}$. In order to suppress fluctuations and achieve convergence we require $\Delta\overline{s}\ll\overline{s}$, i.e. $N_{\rm{tr}}\gg e^{2\gamma_A/\gamma_R}$.

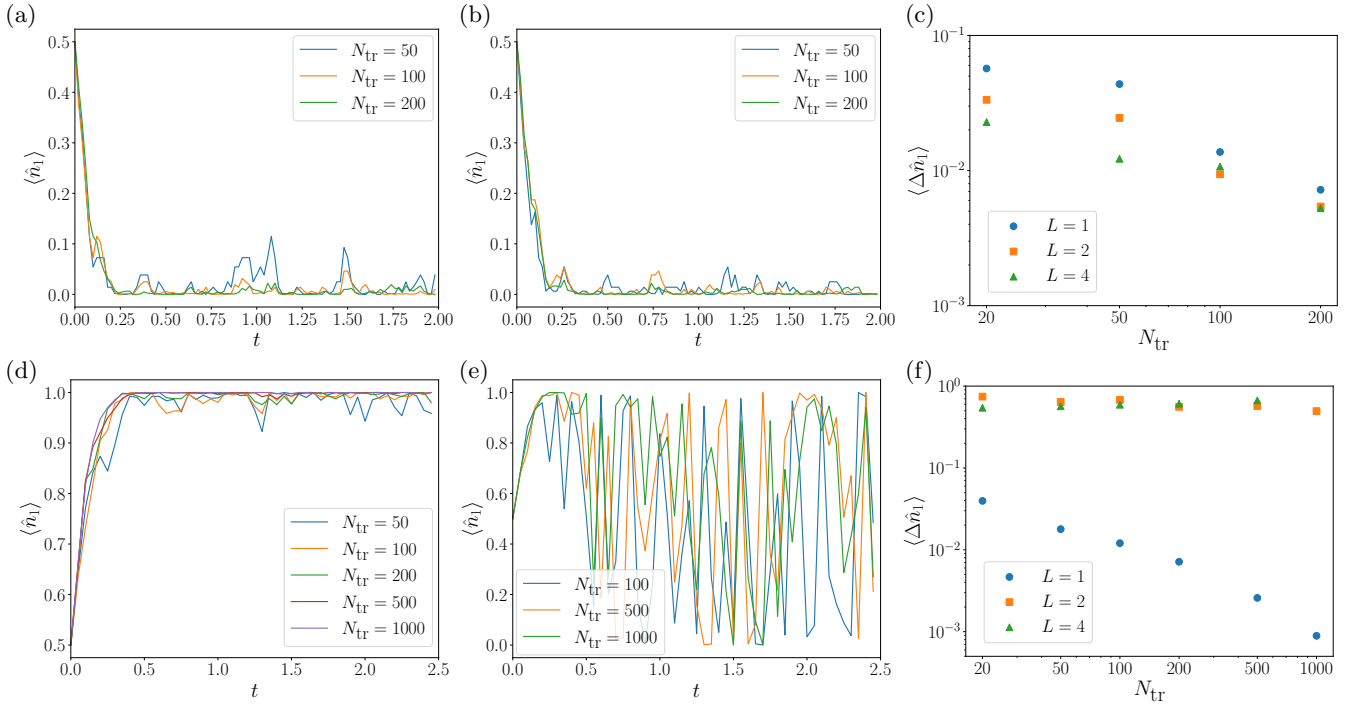
\begin{figure*}
    \setlength{\unitlength}{1cm}
    \includegraphics[width=0.32\linewidth]{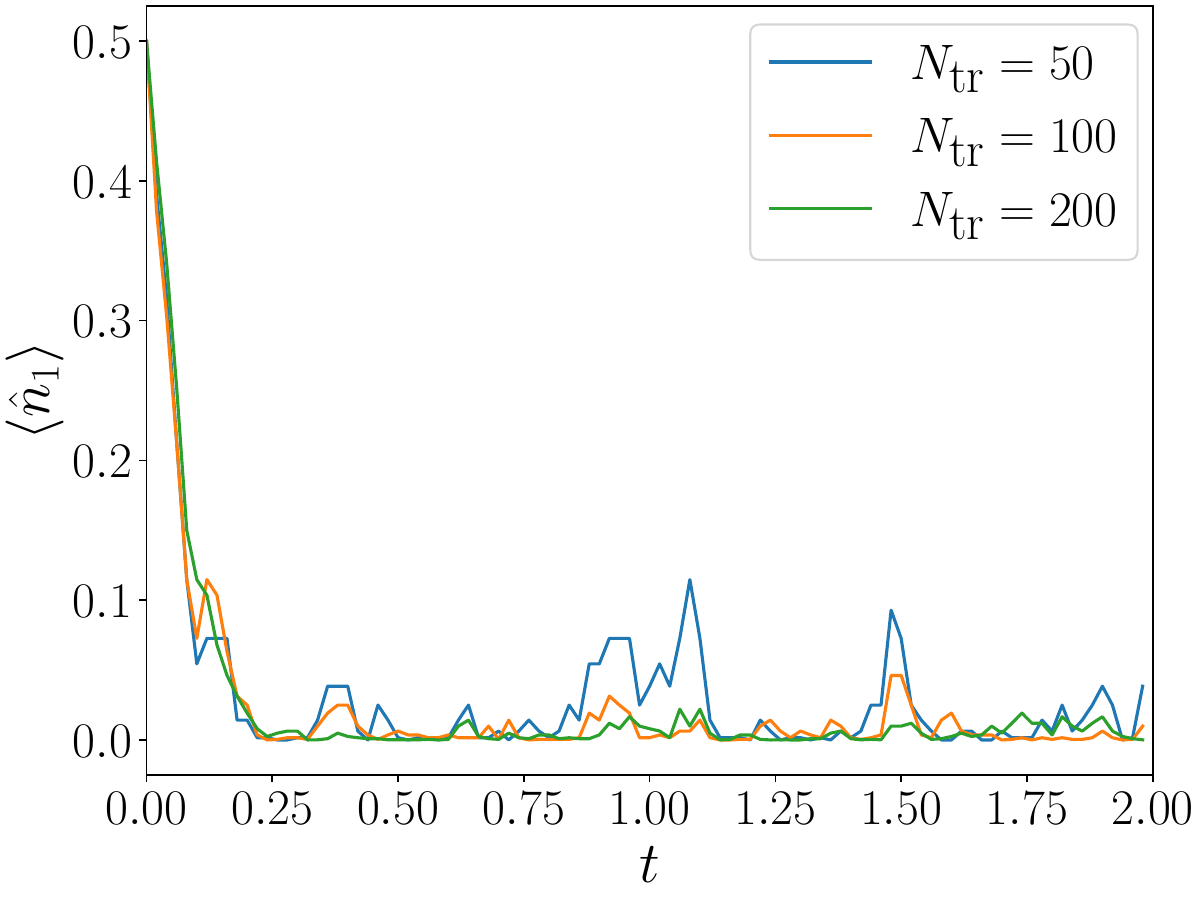}
    \includegraphics[width=0.32\linewidth]{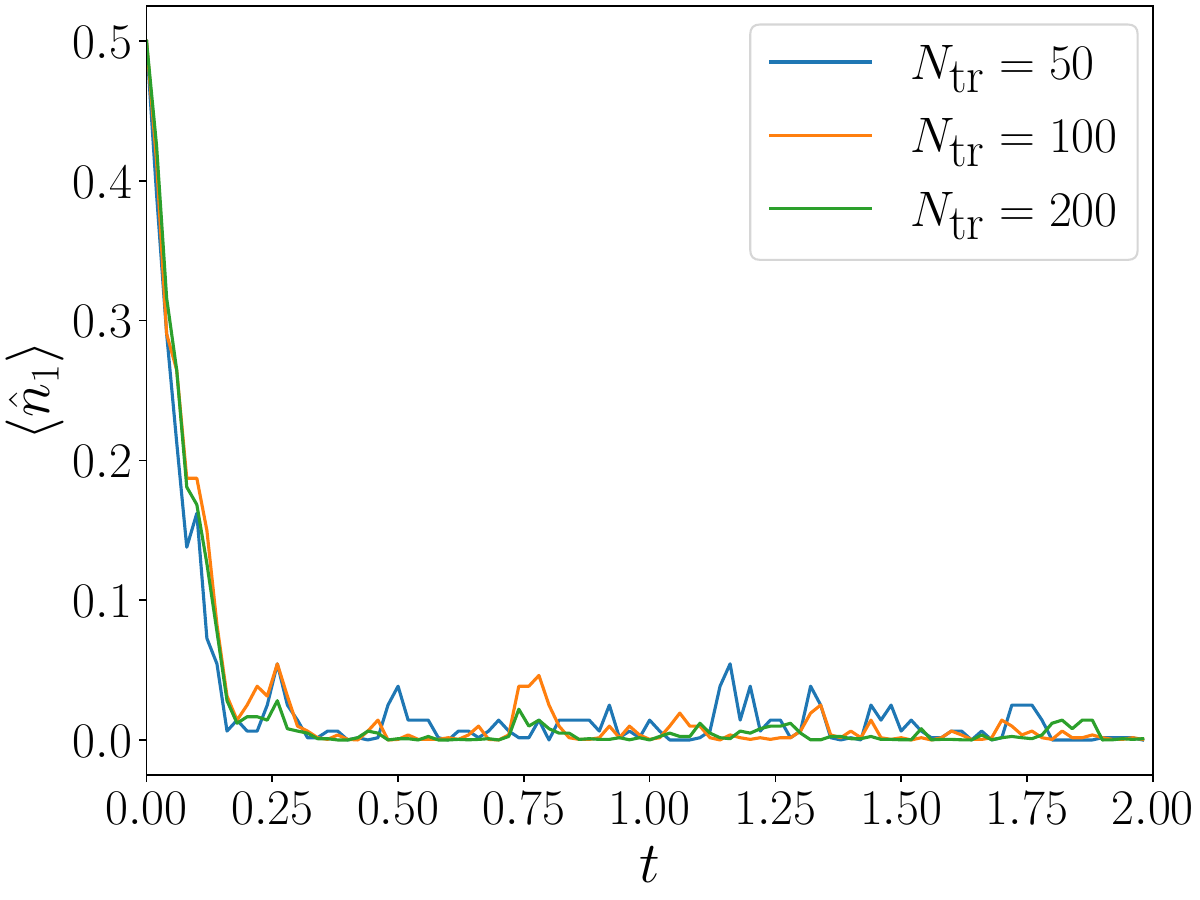}
    \includegraphics[width=0.32\linewidth]{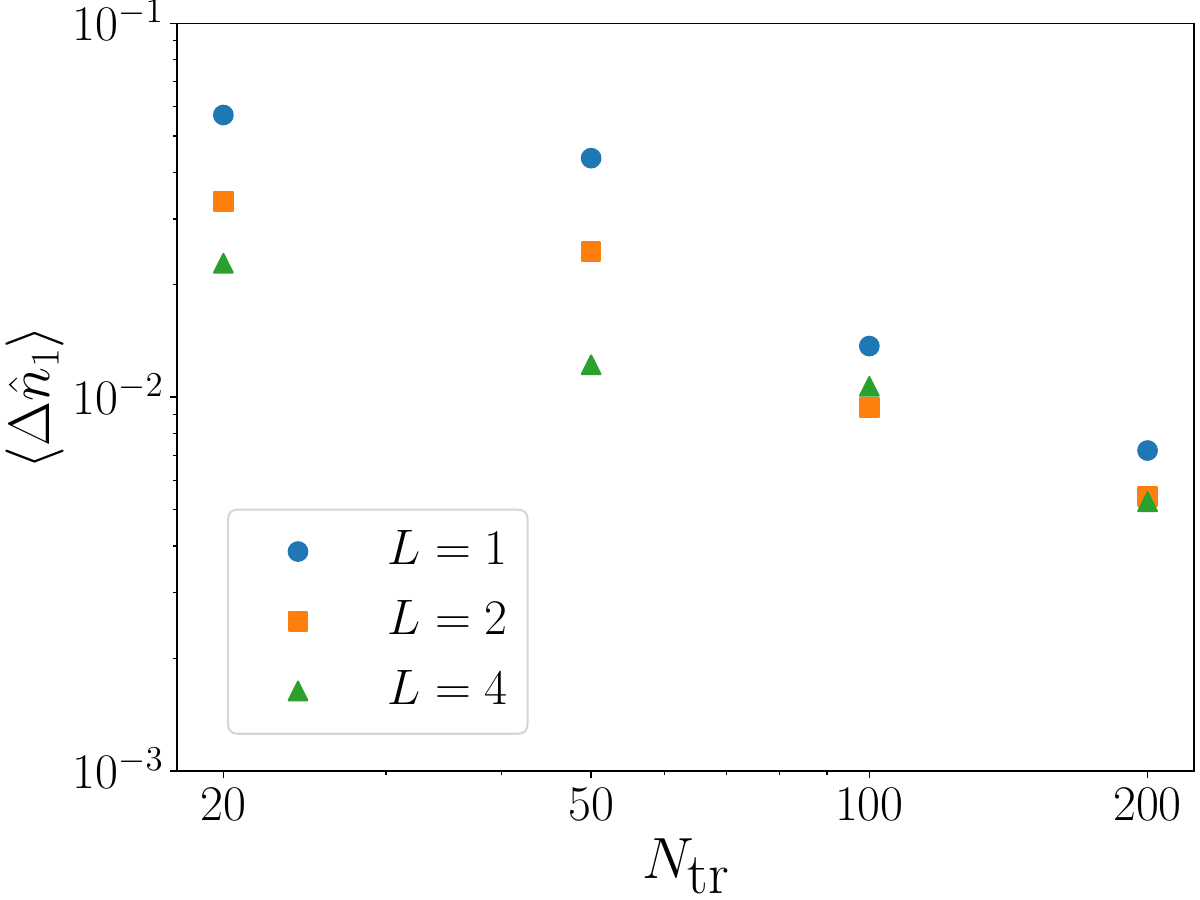}\\
    \vspace*{0.3cm}
    \includegraphics[width=0.32\linewidth]{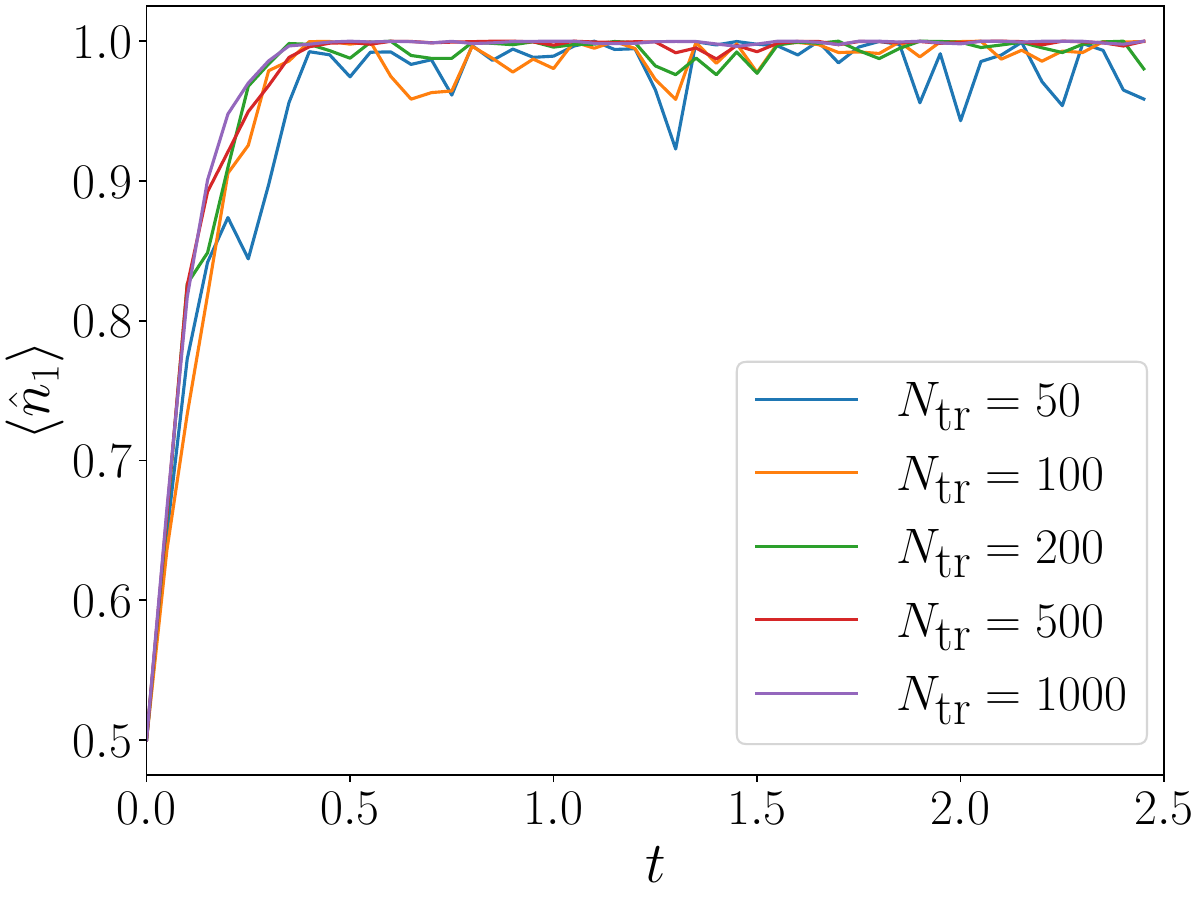}
    \includegraphics[width=0.32\linewidth]{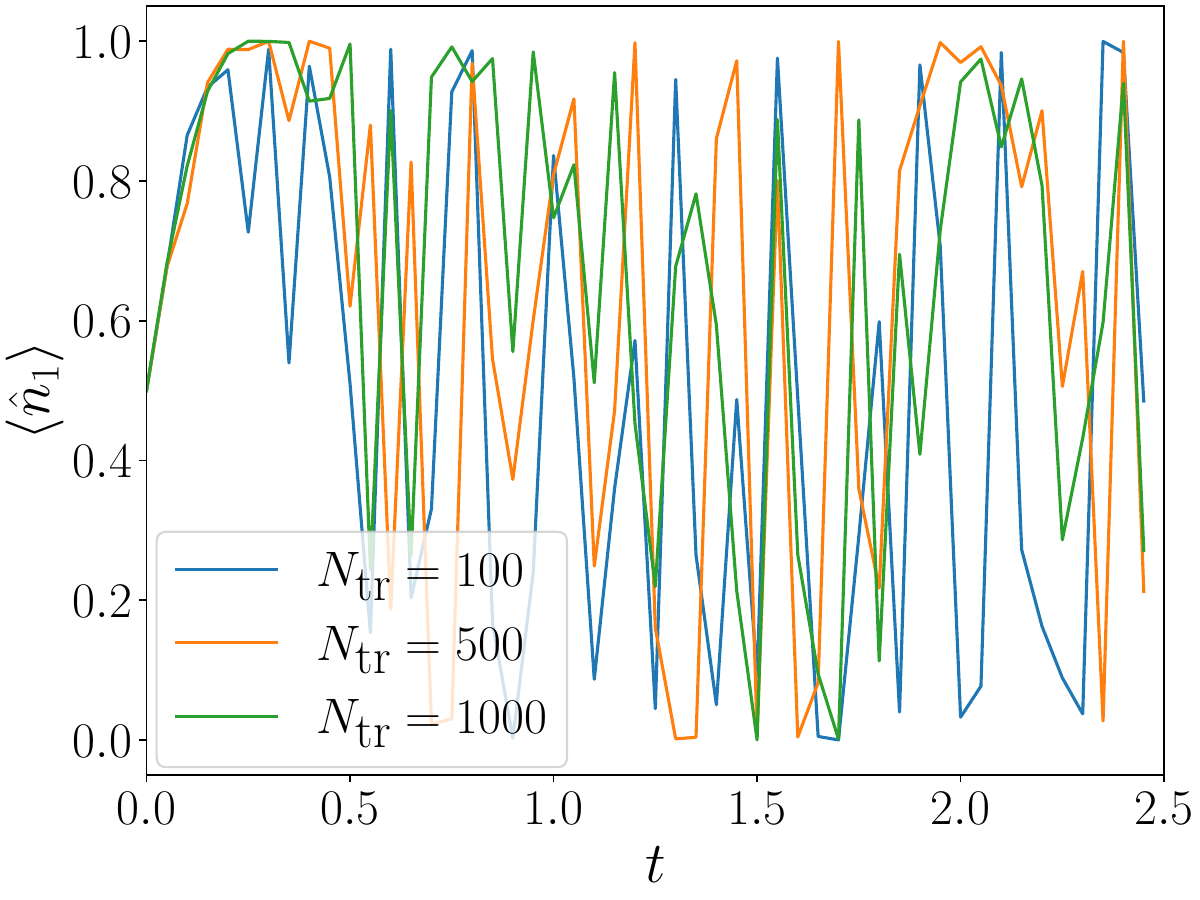}
    \includegraphics[width=0.32\linewidth]{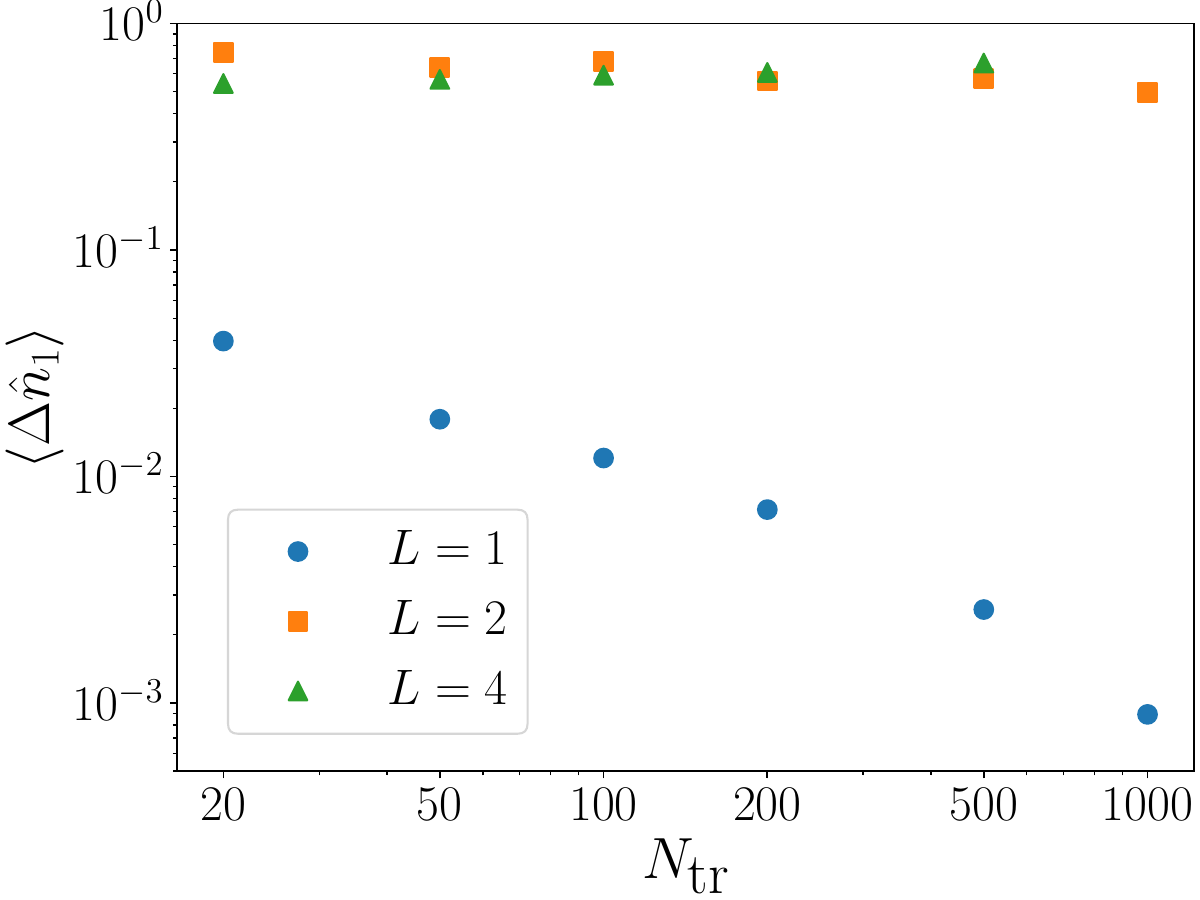}
    \begin{picture}(0,0)
    \put(-17.7,9){(a)}
    \put(-11.7,9){(b)}
    \put(-5.8,9){(c)}
    \put(-17.7,4.3){(d)}
    \put(-11.7,4.3){(e)}
    \put(-5.8,4.3){(f)}
    \end{picture} 
    \caption{Plot of $\langle\hat{n}_{1,\uparrow}(t)\rangle$ as function of the trajectory time $t$. Panels (a)-(b) show the results for the Markovian regime ($\mu=-8$, $J=0$, $U=-4$) for $L=2$ (a) and $L=4$ (b). Panels (d)-(e) show the results for the non-Markovian regime ($\mu=0$, $J=0$, $U=-4$, corresponding to an attractive correlated model) for $L=1$ (d) and $L=2$ (e). Panels (c) and (f) report the dependence of the fluctuations $\Delta n_{1,\uparrow}$ as function of $L$ and $N_{\rm{tr}}$ for the Markovian regime (c) and for the physically realistic, non-Markovian (f) regime.}
    \label{fig:plot}
\end{figure*}

Therefore, the scaling of $N_{\rm{tr}}$ with $L$ depends on how the ratio $\gamma_A/\gamma_R$ scales with $L$. For the Hubbard model (and for any system with a non-zero number of particles), each site $\mbf k$ contributes a chemical potential channel with rate $\gamma_\mu=-\mu/2<0$, so that the classical bit growth rate scales linearly with system size $\gamma_A\sim L\mu$. On the other hand, if we assume that $\gamma_R$ is of the order of single-particle excitation energies (as usually occurs) -- i.e. $\gamma_R\sim\mathcal{O}(\mu,J,U)$ -- we find $\gamma_A/\gamma_R\sim L$. This means that $N_{\rm{tr}}\sim e^L$ and an exponentially large number of trajectories is required to ensure convergence of the non-Markovian unravelling. This means that the complexity is $\mathcal{C}\sim L^2e^L$, i.e. it is still exponential in the size of the system.

It thus seems that the exponential complexity of finding the ground state of a correlated system does not disappear, but rather emerges in the number of trajectories required to accurately simulate the system.

We performed numerical simulations to verify the analytical scaling argument above, and obtained results that confirm the predictions. We studied small system sizes ($L=1,2,4$) for a trivial Markovian case ($U=-4$, $\mu=-8$, $J=0$) and for a physically realistic non-Markovian case ($U=-4$, $\mu=0$, $J=0$). We initialized the system in a state with $\langle\hat n_{1,\uparrow}\rangle=0.5$ and 0 otherwise and let it evolve up to $t=5$. We calculated the expectation value of $\hat n_{1,\uparrow}$ as function of the dynamical time $t$ of the trajectories. The expectation value should converge to 0 for the Markovian case (no particles in the system) and 1 for the non-markovian case (full occupancy). This is what happens in the Markovian case, where the occupancy converges to 0 after a time $t_R\sim0.5$ already for a small number of trajectories $N_{\rm{tr}}\sim20$, see Fig. \ref{fig:plot}(a)-(b). We calculate the typical fluctuation $\Delta n_{1,\uparrow}$, defined as the average over the times after convergence $t\gtrsim t_R$ of $\langle\hat{n}_{1,\uparrow}\rangle^2$:
\begin{equation}\label{eq:fluctuM}
\Delta n_{1,\uparrow}^2\equiv\frac{1}{t_f-t_R}\int_{t_R}^{t_f}dt\langle\hat{n}_{1,\uparrow}(t)\rangle^2.
\end{equation}

We find [Fig. \ref{fig:plot}(c)] that fluctuations decrease by increasing the number of trajectories, and are approximately independent of $L$ when $N_{\rm{tr}}\gtrsim100$, while they slightly decrease with increasing $L$ for small $N_{\rm{tr}}$.

On the other hand, the situation is much different for the non-Markovian case. Even with no hopping terms ($J=0$), we see that convergence of $\langle\hat n_{1,\uparrow}\rangle$ to 1 is worse, and fluctuations are larger. For $L=1$, $\langle\hat n_{1,\uparrow}\rangle$ shows a good convergence behavior, see Fig. \ref{fig:plot}(d); on th other hand, for $L=2$, after a short initial time where convergence is quickly reached, the value of $\langle\hat n_{1,\uparrow}\rangle$ exhibits large fluctuations, see Fig. \ref{fig:plot}(e). In analogy with the Markovian case and Eq. \eqref{eq:fluctuM}, we define the fluctuations also in the non-Markovian case
\begin{equation}\label{eq:fluctuNM}
\Delta n_{1,\uparrow}^2\equiv\frac{1}{t_f-t_R}\int_{t_R}^{t_f}dt(1-\langle\hat{n}_{1,\uparrow}(t)\rangle)^2,
\end{equation}
since we care about deviations from the NESS value $\langle\hat{n}_{1,\uparrow}(t)\rangle=1$. In Fig. \ref{fig:plot}(f), we plot the fluctuations as function of $L$ and $N_{\rm{tr}}$. While for $L=1$ we observe a dependence on $N_{\rm{tr}}$ similar to the Markovian case, already for $L=2$ and $L=4$ fluctuations are much larger (of order 1) and do not show a significant decay with larger $N_{\rm{tr}}$. 

The results of these simulations show that unravelling the non-Markovian dynamics requires a large number of trajectories which is highly dependent on the size of the system, contrary to the Markovian case, so that the computational cost remains exponential in system size.

\section{Conclusions}\label{Sec:Conclusions}

In this manuscript, we have introduced a mapping between the ground state of a system of correlated electrons and an open system of Majorana fermions with a purely dissipative dynamics.

We have established the rules to map a generic Hamiltonian into a Lindbladian, including how to determine the jump operators and the dissipative rates. We have found that both single-body terms and two-body interactions map into jump operators that preserve the Gaussianity of the Majorana fermions. This property can be exploited by unravelling the dissipative dynamics in terms of quantum trajectories, where each trajectory preserves Gaussian states and can thus be characterized only via the correlation matrix, making the numerical simulations of a trajectory viable even for large system sizes.

We find that correlated systems with a non-zero number of particles map onto non-Markovian dynamics. Such dynamics can be unravelled via an auxiliary classical bit, but the number of trajectories needed to achieve convergence dramatically increases, exhibiting an exponential dependence with the size of the system. Thus while the simulation of a single trajectory is numerically easy, the large number of trajectories required makes the computational complexity of the problem still exponential.

In summary, the mapping developed in this work provides a new perspective on the relationship between strongly correlated quantum matter and open quantum systems. While the trajectory formulation does not eliminate the exponential complexity associated with finding correlated ground states, it reveals how this complexity is transferred from the representation of the quantum state to the statistical sampling required by the non-Markovian trajectory unravelling. In particular, individual trajectories remain Gaussian and can be simulated efficiently using only two-point correlation functions, whereas the computational bottleneck emerges due to the exponentially large number of trajectories required to overcome the sign problem associated with negative dissipative rates.

Beyond its conceptual implications, this framework opens several directions for future research. The new perspective it provides may prove useful in the search for alternative unravellings of non-Markovian dynamics, aimed at decreasing fluctuations, or solutions inspired by sign-problem free models in quantum Monte Carlo that reduce the trajectory complexity. More broadly, the mapping established here suggests that tools developed in open quantum systems and non-Markovian dynamics may provide new insights into longstanding computational challenges in many-body physics.

\acknowledgments{We thank Marcello Dalmonte and Rosario Fazio for insightful discussions.

This work was supported by ICSC – Centro Nazionale di Ricerca in High-Performance Computing, Big Data and Quantum Computing under project E63C22001000006. G.C. acknowledges the CINECA award under the ISCRA initiative, for the availability of high performance computing resources and support.}

\bibliography{FH}

\clearpage

\onecolumngrid
\appendix

\section{Mapping rules}\label{app:Mapping}

Let us assume that the density matrix of the system is described by $\rho=\prod_{\mbf k}w_{\mbf k}^{\alpha_{\mbf k}}$. Then the following identities hold \cite{prosenThirdQuantizationGeneral2008,seligmanThirdQuantization2010}
\begin{equation}\label{eq:rhowk}
\rho w_{\mbf k}=(-1)^{\alpha+\alpha_{\mbf k}}w_{\mbf k}\rho;\qquad \alpha_{\mbf k}\rho =(-1)^{\alpha}w_{\mbf k}\rho w_{\mbf k};
\end{equation}
where $\alpha\equiv\sum_\mbf k\alpha_\mbf k$.

The rules for mapping operators are the following

\begin{gather}
    \hat c^\dagger_{\mbf k}\ket{\psi_\rho}\leftrightarrow\frac12(1+(-1)^{\alpha_{\mbf k}})w_{\mbf k}\rho;\qquad     \hat c_{\mbf k}\ket{\psi_\rho}\leftrightarrow\frac12(1-(-1)^{\alpha_{\mbf k}})w_{\mbf k}\rho;\\
    \hat c^\dagger_{\mbf k}\hat c_{\mbf k}\ket{\psi_\rho}\,\,\leftrightarrow\,\,\frac12(1+(-1)^{\alpha_{\mbf k}+1})w_\mbf k\frac12(1-(-1)^{\alpha_{\mbf k}})w_\mbf k\rho=\frac12(1+(-1)^{\alpha_{\mbf k}})\rho=\alpha_{\mbf k}\rho=\frac12(\rho-(-1)^\alpha w_{\mbf k}\rho w_{\mbf k});\\
    (1-2\hat n_{\mbf k})\ket{\psi_\rho}\,\,\leftrightarrow\,\,(-1)^\alpha w_{\mbf k}\rho w_{\mbf k};
\end{gather}

For systems with an even number of fermions, $\alpha$ is even and thus $w_{\mbf k}\rho w_{\mbf k}\,\,\leftrightarrow\,\,(1-2\hat n_{\mbf k})\ket{\psi_\rho}$

\begin{gather}
    \mathcal{L}[w_{\mbf k}]\rho=w_{\mbf k}\rho w_{\mbf k}-\frac12\{w_{\mbf k}^2,\rho\}=w_{\mbf k}\rho w_{\mbf k}-\rho\,\,\leftrightarrow\,\,-2\hat n_{\mbf k}\ket{\psi_\rho};\\
    w_{\mbf k}\rho w_{\mbf l}-\frac12\{w_{\mbf l}w_{\mbf k},\rho \}\,\,\leftrightarrow\,\,(2\hat c^\dagger_{\mbf k}\hat c_{\mbf l}-\hat c^\dagger_{\mbf k}\hat c_{\mbf l}-\hat c^\dagger_{\mbf l}\hat c_{\mbf k})\ket{\psi_\rho};\\
    \mathcal{L}\left[\frac{w_{\mbf k}+w_{\mbf l}}{\sqrt2}\right]\rho\,\,\leftrightarrow\,\,-(\hat c^\dagger_{\mbf k}+\hat c^\dagger_{\mbf l})(\hat c_{\mbf k}+\hat c_{\mbf l})\ket{\psi_\rho};\\
    \mathcal{L}[w_{\mbf k}w_{\mbf l}]\rho=w_{\mbf k}w_{\mbf l}\rho w_{\mbf l}w_{\mbf k}-\rho\,\,\leftrightarrow\,\,[(2\hat n_{\mbf k}-1)(2\hat n_{\mbf l}-1)-1]\ket{\psi_\rho}=[4\hat n_{\mbf k}\hat n_{\mbf l}-2(\hat n_{\mbf k}+\hat n_{\mbf l})]\ket{\psi_\rho};
\end{gather}

\section{Update of the correlation matrix}\label{app:UpdateCorr}

The effect of the dissipative dynamics on the correlation matrix along a single quantum trajectory can be easily calculated. If the system performs no jumps, then $A_{\bf k\bf l}$ remains unchanged. On the other hand, when the system performs a jump with operator $w_{\mbf j}$, the density matrix changes according to
\begin{gather}
    \rho'=w_{\mbf j}\rho w_{\mbf j}=w_{\mbf j}e^{\frac14\sum_{\mbf k\mbf l}A_{\mbf k\mbf l}w_{\mbf k}w_{\mbf l}}w_{\mbf j}=e^{\frac14\sum_{\mbf k\mbf l}A_{\mbf k\mbf l}w_{\mbf j}w_{\mbf k}w_{\mbf l}w_{\mbf j}}=e^{\frac14\sum_{\mbf k\mbf l}A'_{\mbf k\mbf l}w_{\mbf k}w_{\mbf l}}
\end{gather}
where we have used the property that $w_{\mbf j}$ is unitary since $w_{\mbf j}=w_{\mbf j}^\dagger$ and $w_{\mbf j}^2=1$. Using the anticommutation rules for the Majorana operators we find
\begin{equation}
w_{\mbf j}w_{\mbf k}w_{\mbf l}w_{\mbf j}=w_{\mbf k}w_{\mbf l}-2w_{\mbf k}w_{\mbf l}\delta_{\mbf k\mbf j}-2w_{\mbf k}w_{\mbf l}\delta_{\mbf l\mbf j}+4w_{\mbf k}w_{\mbf l}\delta_{\mbf k\mbf j}\delta_{\mbf l\mbf j};\qquad A_{\mbf k\mbf l}w_{\mbf j}w_{\mbf k}w_{\mbf l}w_{\mbf j}=A'_{\mbf k\mbf l}w_{\mbf k}w_{\mbf l},
\end{equation}
where $A'_{\mbf k\mbf j}=-A_{\mbf k\mbf j}$, $A'_{\mbf j\mbf l}=-A_{\mbf j\mbf l}$, $A'_{\mbf j\mbf j}=A_{\mbf j\mbf j}$ and all the other elements stay the same. In other words, applying $w_{\mbf j}$ multiplies the $\mbf j$-th row and the $\mbf j$-th column of $A$ by $-1$, with the understanding that the $\mbf j$-$\mbf j$ element remains the same.

Similarly, applying $L_\nu=\frac{w_{\mbf i}+w_{\mbf j}}{\sqrt2}$ gives
\begin{gather}
    A'_{\bf k\bf i}=-A_{\bf k\bf j};\qquad A'_{\bf k\bf j}=-A_{\bf k\bf i}; \qquad A'_{\bf i\bf l}=-A_{\bf j\bf l}; \qquad A'_{\bf j\bf l}=-A_{\bf i\bf l};\qquad A'_{\mbf i\mbf j}=A_{\mbf i\mbf j}
\end{gather}
In other words the action of any jump operator affects only the rows and columns of $A$ corresponding to the Majorana fermions in the jump operator. Thus, the computational complexity of implementing a quantum jump is $\mathcal{O}(L)$ instead of $\mathcal{O}(L^2)$ since only a few rows instead of the entire matrix need to be updated.

\section{Evolution of the density matrix and classical bit}\label{App:BitEvol}

The dynamical equations for averages over trajectories of the density matrix and of the classical bit can be found starting from their update rules and then averaging over trajectories. The average is equivalent to averaging over the probabilities of the various jumps or of the non-hermitian dynamics during a single time step $dt$:
\begin{gather}
s_a(t+dt)=\begin{cases}
  s_a(t) & \text{  with probability  } \sum_{\gamma_\nu>0}\gamma_\nu dt \\
  -s_a(t) & \text{  with probability  } \sum_{\gamma_\nu<0}|\gamma_\nu| dt \\
  \frac{1-\sum_\nu\gamma_\nu dt}{1-\sum_\nu|\gamma_\nu| dt}s_a(t) & \text{  with probability  } 1-\sum_\nu|\gamma_\nu|dt
\end{cases}\\
\label{eq:stdt}\overline{s(t+dt)}=\sum_{\gamma_\nu>0}\gamma_\nu dt\overline{s(t)}-\sum_{\gamma_\nu<0}|\gamma_\nu| dt\overline{s(t)}+(1-\sum_\nu|\gamma_\nu| dt)\frac{1-\sum_\nu\gamma_\nu dt}{1-\sum_\nu|\gamma_\nu| dt}\overline{s(t)}=\overline{s(t)}
\end{gather}
where $\overline{s}$ is the average over trajectories of $s$. The equality implies that $\overline{s(t)}$ is constant in time and thus equal to 1, its initial value.

The average of the density matrix is
\begin{gather}
s_a(t+dt)\rho_a(t+dt)=\begin{cases}
  \text{sign}(\gamma_\nu)s_a(t)L_\nu\rho_a(t)L^\dagger_\nu & \text{  with probability  } |\gamma_\nu| dt \\
  \frac{1-\sum_\nu\gamma_\nu dt}{1-\sum_\nu|\gamma_\nu| dt}s_a(t)\rho_a(t) & \text{  with probability  } 1-\sum_\nu|\gamma_\nu|dt
\end{cases}\\
\notag\rho(t+dt)=\sum_{\nu}|\gamma_\nu| dt\text{ sign}(\gamma_\nu)\overline{s_a(t)L_\nu\rho_a(t)L^\dagger_\nu}+(1-\sum_\nu|\gamma_\nu| dt)\frac{1-\sum_\nu\gamma_\nu dt}{1-\sum_\nu|\gamma_\nu| dt}\overline{s_a(t)\rho_a(t)}\\
\label{eq:rhotdt}\rho(t+dt)=\sum_\nu\gamma_\nu dtL_\nu\rho(t)L^\dagger_\nu+(1-\sum_\nu\gamma_\nu dt)\rho(t)=\rho(t)+dt\sum_\nu\gamma_\nu\left[L_\nu\rho(t)L^\dagger_\nu-\frac12\{L^\dagger_\nu L_\nu,\rho(t)\}\right]
\end{gather}
where we have used that $L^\dagger_\nu L_\nu=\mathbb{1}$.

For the average of the square of the classical bit, required to calculate fluctuations, we find
\begin{gather}
s_a^2(t+dt)=\begin{cases}
  s^2_a(t) & \text{  with probability  } \sum_\nu|\gamma_\nu|dt \\
  \left(\frac{1-\sum_\nu\gamma_\nu dt}{1-\sum_\nu|\gamma_\nu| dt}\right)^2s^2_a(t) & \text{  with probability  } 1-\sum_\nu|\gamma_\nu|dt
\end{cases}\\
    \overline{s^2(t+dt)}=\sum_\nu|\gamma_\nu|dt\overline{s^2(t)}+\frac{(1-\sum_\nu\gamma_\nu dt)^2}{1-\sum_\nu|\gamma_\nu| dt}\overline{s^2(t)}\approx\overline{s^2(t)}[1+2\sum_{\nu}(|\gamma_\nu|-\gamma_\nu)dt]=\overline{s^2(t)}[1+2\sum_{\gamma_\nu<0}|\gamma_\nu|dt]
\end{gather}

From Eq. \eqref{eq:stdt} we can also prove that $\text{Tr}\rho(t)=1$. In fact,
\begin{equation}
    \text{Tr}\rho(t)=\frac{1}{N_{\rm{tr}}}\sum_{a=1}^{N_{\rm{tr}}}s_a(t)\text{Tr}\rho_a(t)=\frac{1}{N_{\rm{tr}}}\sum_{a=1}^{N_{\rm{tr}}}s_a(t)=\overline{s(t)}=1
\end{equation}

\end{document}